\documentclass{article}

\usepackage{PRIMEarxiv}
\usepackage{tabto}
\usepackage[utf8]{inputenc} % allow utf-8 input
\usepackage[T1]{fontenc}    % use 8-bit T1 fonts
\usepackage{hyperref}       % hyperlinks
\usepackage{url}            % simple URL typesetting
\usepackage{booktabs}       % professional-quality tables
\usepackage{amsfonts}       % blackboard math symbols
\usepackage{nicefrac}       % compact symbols for 1/2, etc.
\usepackage{microtype}      % microtypography
\usepackage{lipsum}
\usepackage{fancyhdr}       % header
\usepackage{graphicx}       % graphics
\graphicspath{{figures/}}     % organize your images and other figures under media/ folder
\usepackage{gensymb}
\usepackage{amsmath}% http://ctan.org/pkg/amsmath

\title{ Analysis of Synchrosqueezed Transforms and Application Perspectives
}

\author{
  Rayyan Abdalla \\
  Arizona State University \\
  \texttt{rsabdall@asu.edu} \\
}

\begin{document}
\maketitle

\begin{abstract}
\hspace{.3in}High-resolution time-frequency (TF) analysis plays crucial role in characterizing
multicomponent signal (MCSs) and estimating oscillatory properties. Linear time-frequency
representations (TFRs) such as classical short-time Fourier transform (STFT) and continuous wavelet
transform (CWT) incur constrained TF resolution and energy diffusion in both time and frequency direction.
The synchrosqueezing transform (SST) represents a powerful sparse reassignment method that allows
component reconstruction. This work introduces SST as extension to STFT and CWT and illustrates
corresponding advantages of sharpening TFRs and recovery of instantaneous components. The SST
effectiveness is assessed in practical situations that involve comparing STFT-based and CWT-based
versions of synthetic data and also applying SST to optimize deep learning (DL) prediction model. It is
demonstrated how SST achieves promising results in terms of improving TFR readability and increasing
accuracy of DL-based prediction models.
\end{abstract}

\section{Introduction}
 \hspace{.3in}The class of non-stationary signals encompasses signals with time-varying oscillatory components and hence arises essentially in numerous scientific and physical world applications. Harmonic analysis and
processing of a non-stationary signal extends beyond the traditional Fourier transform in order to highlight
the signal frequency information with varying amplitudes in a time-localized manner.\par
\hspace{.3in}Standard TF analysis involves linear TFRs among which the most popular are STFT and CWT.
While both STFT and CWT provide invertible local transformations, both TFRs are constrained by TF
resolution trade-off according to Heisenberg–Gabor uncertainty principle \cite{thakur2015synchrosqueezing}. The Wigner Distribution
(WD), which is a quadratic TFR, was proposed in attempt to tackle TF resolution limitation. However, WD
exhibits interference in the form of cross-terms between different signal components and thence hampersreadability and extraction of instantaneous properties. As regards invertibility and component retrieval of
an MSC, the empirical mode decomposition (EMD) algorithm was introduces for MCSs decomposition.
Nonetheless, EMD lacks firm mathematical foundation and basically experiences the problem of mode
mixing \cite{li2012generalized}. \par
\hspace{.3in}The SST proposes an enhanced reassignment approach that adapts to linear TFRs. Particularly, SST
procedure applies post-processing to both STFT and CWT in order to concentrate energy closer to most
prominent frequencies. While SST is still limited by uncertainty principle, the signal spectrum is condensed
in the frequency direction and thus resulting in an improved TF resolution. Moreover, SST is sparse and
invertible in a sense that it offers exact reconstruction formula that allow mode reproducibility of MSCs
modes \cite{thakur2015synchrosqueezing}. This work comprehensively reviews SST theory and individual mode reconstruction. We also
discuss and SST suitability for real-life applications, mainly TF analysis associated to EEG seizure
prediction. We compare and contrast conventional STFT and CWT with their synchrosqueezed versions
and demonstrate how SST provides various texture features for DL classification approaches.\par
\hspace{.3in}In this work, we provide a concise illustration of SST methodology in section 2. Both STFT and
CWT versions are discussed with their mode reconstruction formulas. Section 3 introduces practical
implementation of SST. The relative performance of STFT-based SST and CWT-based SST is discussed
using a synthetic example. The SST is then implemented to optimize performance of DL-based epileptic
seizure prediction system. Results and outcomes are discussed in section 4 and conclusions are drawn in
section 5.

\section{The Synchrosqueezing Theory}
\label{sec:system model} 
 \hspace{.3in}We consider analyzing an MSC with $N$ time-varying harmonic components/modes represented by
equation \ref{eq1}
 \begin{equation} \label{eq1} 
     x(t) = \sum_{k=1}^{N} A_k(t)e^{j2\pi \phi_k(t)} \
 \end{equation}
Where $A_k(t)$ is the instantaneous amplitude and $\phi_k(t)$ is the instantaneous phase function of a
component $k$, respectively. The SST process mainly pursues sharpening linear TFRs, i.e. STFT and CWT,
and establishes mode synthesis formula to retrieve time-varying modes. We hereafter present both CWT
and STFT based settings as developed in \cite{thakur2015synchrosqueezing} and \cite{meignen2019synchrosqueezing}.

\subsection{STFT-based SST}
\hspace{.3in}Assume $S^h_x(t,f)$ to be the STFT of a signal $x(t)$ taken by sliding a window $h(t)$, the STFT-based SST
is given by equation \ref{eq2}.
\begin{equation}  \label{eq2}
    Tsx^{\varepsilon , M}_{\gamma} (t,\eta) = \int_{ \{ (t,f): f \in [M^{-1},M], |S_x(t,f)|>\gamma \} } S^h_x(t,f) \frac{1}{\varepsilon} g \left( \frac{\eta - \hat{\omega}_x^S (t,f)}{\varepsilon}\right) \,df \
\end{equation}

Where $\gamma$ is an energy threshold specified such that STFT coefficients with energy larger than $\gamma$ are
squeezed in the frequency domain. The frequency scope of the processed STFT is defined by parameter $M$.
The factor $\varepsilon$ compensates for errors in calculating  $\hat{\omega}_x^S (t,f)$ which represents instantaneous frequency
estimate and is derived by phase transform equation given in \ref{eq3}
\begin{equation} \label{eq3}
    \hat{\omega}_x^S (t,f) = \frac{\frac{\partial}{\partial t} S^h_x(t,f)}{2 \pi i S^h_x(t,f)}
\end{equation}

\hspace{0.3in}The $g$ function represents a test function which is a practical approximation of $\sigma \left(  \eta - \hat{\omega}_x^S (t,f) \right)$. \par
That is to say, the STFT-based SST attempts to map the STFT $(t,f)$ plane to a new plane $(t,\eta)$ around the approximated instantaneous frequencies. \par
Any component $k$ can be reconstructed by inverting SST formula as shown in equation \ref{eq4}

\begin{equation} \label{eq4}
    Rs_{k,\gamma}^{\varepsilon, M} (t) = \frac{1}{\int_{- \infty}^{\infty} |h(z)|^2 dz} \int_{|\eta - \phi_k'| < d} Tsx^{\varepsilon , M}_{\gamma} (t,\eta) d \eta 
\end{equation}
Where $d$ represents a compensation error to avoid mode mixing when estimating the actual $k^{th}$
instantaneous frequency $\phi_k'$. \par

\subsection{CWT-based SST}
\hspace{0.3in}The CWT setting is similar to that of STFT with slightly different assumptions. Consider $W^{\psi}_x(a,t)$ to
be the CWT of a signal $x(t)$ in terms of a mother wavelet $\psi (t)$, the CWT-based SST is given by equation \ref{eq5}
\begin{equation}  \label{eq5}
    Twx^{\varepsilon , M}_{\gamma} (t,\eta) = \int_{ \{ (a,t): a \in [M^{-1},M], |S_x(a,t)|>\gamma \} } W^{\psi}_x(a,t) \frac{1}{\varepsilon} g \left( \frac{\eta - \hat{\omega}_x^S (a,t)}{\varepsilon}\right) \,df \
\end{equation}

\hspace{0.3in} All parameters definitions hold as prescribed in STFT-based SST, except that the parameter $M$ here
represents the CWT scale scope within which CWT is processed. The phase transform in the context of
CWT is given by equation \ref{6}
\begin{equation} \label{eq6}
    \hat{\omega}_x^S (a,t) = \frac{\frac{\partial}{\partial t} W^{\psi}_x(a,t)}{2 \pi i W^{\psi}_x(a,t)}
\end{equation}
Subsequently, the $k^{th}$ mode can be retrieved using the formula of equation \ref{eq7}.
\begin{equation} \label{eq7}
    Rs_{k,\gamma}^{\varepsilon, M} (t) = \frac{1}{\int_{0}^{\infty} \frac{\psi (z)}{z} dz} \int_{|\eta - \phi_k'| < d} Twx^{\varepsilon , M}_{\gamma} (t,\eta) d \eta 
\end{equation}

It is important to note that the approximation $Rs_{k,\gamma}^{\varepsilon, M} (t) 	\approx A_k(t)e^{j2\pi \phi_k(t)}$ is possible under certain 
conditions depending on the ridge extraction algorithm used and how harmonically separated the modes of
the recovered MCS \cite{meignen2019synchrosqueezing}.

\section{Practical Implementation}
\label{sec:section4}
\hspace{.3in}In practice, the performance of SST implementation depends on two main parameters: the selected
window or wavelet in the definition of the linear TFR and the energy threshold in the definition of the SST
formula. Moreover, the choice of the linear TFR to be squeezed impacts directly the resultant TF resolution
\cite{meignen2019synchrosqueezing}. We hereafter inspect the performance of SST using synthetic example and later apply SST to a real-life
data to demonstrate application potentials. Both synthetic and real data SST tests are executed using open-source synchrosqueezing Python tool developed by \cite{muradeli_2022}

\subsection{Analysis of a synthetic signal}
\hspace{0.3in}The SST is tested with an MSC of 3 oscillatory components using both STFT-based and CWT-based
settings. The inspected signal comprises two signals of polynomial instantaneous frequencies and a third
signal of a linear chirp. The signal length is 10 seconds and is sampled by sampling frequency of 205 Hz.
As regards parameter selection, STFT-based setting applies Slepian window of length 32 sample points (i.e.
0.3125 seconds). Concerning CWT-based setting, the wavelet selected is generalized Morse wavelet
(GMW) with a symmetry parameter 3 and decay/compactness parameter 60.

\begin{figure}[!b]
    \centering
    \includegraphics[scale=1]{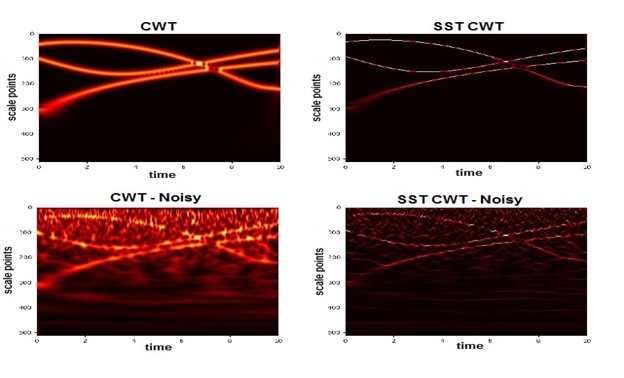}
    \caption{Illustration of 3-component signal represented by CWT and their SST version in
clean/noisy settings}
    \label{fig1}
\end{figure}

\begin{figure}[!tb]
    \centering
    \includegraphics[scale=1]{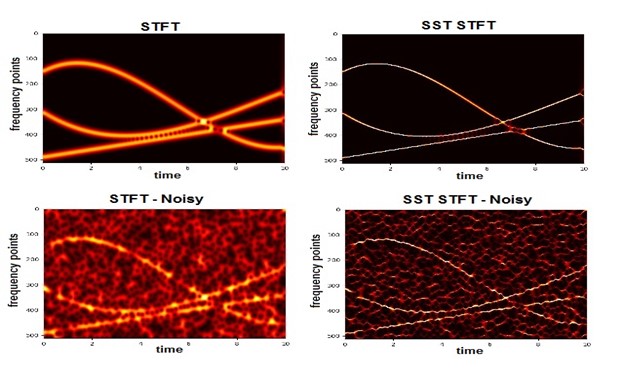}
    \caption{Illustration of 3-component signal represented by STFT and their SST version in
clean/noisy settings}
    \label{fig2}
\end{figure}

\hspace{0.3in} Figures \ref{fig1}  and \ref{fig2} illustrate original/noised CWT/STFT and their synchrosqueezed versions of the inspected
MSC. Following the observation of [1, 3, 4], both STFT and CWT smear the energy around the center
frequency which may lead to mode mixing when analyzing an MCS of closely-packed instantaneous
frequencies. On the other hand, the SST exhibits much improved readability in frequency direction. In other
words, for both STFT-based and CWT-based settings, the frequency resolution is optimized while time
resolution remains unchanged, therefore improved detectability in noisy environments. As the CWT
provides good frequency resolution at lower frequencies, the CWT-based SST gets sharper with the
increasing frequency, unlike STFT-based SST, which demonstrates constant resolution over all frequencies.
One final note suggests that STFT-based STFT is suitable for decomposing MCSs with closely modulated
instantaneous frequencies as CWT-based SST is more convenient for studying lower-frequencies.

\subsection{Application to Prediction of Epilepsy Seizures}

\subsubsection{Overview}
\hspace{0.3in} Epilepsy is a neurological disorder that affects the central nervous system causing abnormal brain
activities and seizures of unusual behaviors. Epilepsy control and management essentially requires early
prediction of seizure occurrence in order to provide timely treatment for patient. As seizures are closely
related to brain synchronization patterns, epileptic patient states are investigated using
electroencephalogram (EEG) signals. However, EEG signals analysis is a complex random process as EEG
patterns are hardly discernible. Moreover, EEG data varies uncertainly with patient normal activities (interpatient variability) and from one patient to another (intra-patient variability). While classical seizure prediction systems involved intensive statistical computations to extract meaningful features from EEG
patterns, yet these statistical method do not provide a generalized theoretical foundation for detecting
seizure onset. Recently, various automatic prediction systems consider employing deep learning (DL)
techniques to identify feature patterns related to events of seizures. The problem statement addresses
discerning between states prior to seizure onset, i.e. pre-ictal states, and states of regular patient brain tasks,
i.e. inter-ictal states \cite{hussein2020augmenting}.

\begin{figure}[!htb]
    \centering
    \includegraphics[width=15cm,height=6cm]{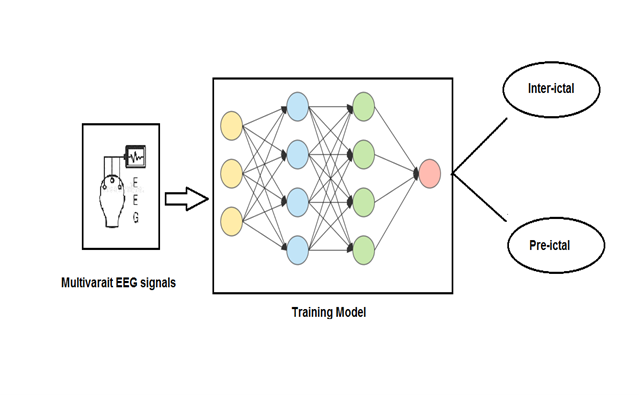}
    \caption{Abstract diagram showing the input (EEG signal) and the output (inter-ictal or pre-ictal) of DL-based seizure prediction model}
    \label{fig3}
\end{figure}

\hspace{0.3in}Potential DL approaches suggest the use of convolutional neural networks (CNN) models with
multiple convolution layers to learn information from raw EEG data. Time-series EEG signals are converted
to two dimensional images (2D) in order to extract deep features from the data. As EEG signals are
extremely noisy and represent huge amount of information, a typical EEG signal contains multiple
oscillatory time-varying components that can be modeled as an MSC. Therefore, imagery information of
an EEG signal can be parsed using high-resolution TF analysis techniques to highlight possible texture
features related to both pre-ictal and inter-ictal states. Figure \ref{fig3} illustrates a DL-based seizure prediction
model using analyzing TFRs obtained from multivariate EEG data.

\hspace{0.3in}While conventional DL-based prediction models utilize linear TFRs, such as STFT and CWT, to
convert EEG data to featured images, we address a practical example that considers applying SST analysis
to preprocess raw EEG data for the purpose of feeding the resultant TFRs to a CNN based seizure detection
model. The following subsections describe the implementation process.

\subsubsection{Dataset description and preprocessing}
\hspace{0.3in}As of any DL-based training model, data quality and versatility is crucial as it directly affects model
accuracy. The random nature of seizure disorders manifests mining datasets of different patients, recorded
at different time segments and accounting for various neurological states. For this purpose, the adopted
dataset represents intracranial EEG recordings of patients and canine subjects which is made available by
American epilepsy society \cite{kaggle}. The training data clips are categorized into pre-ictal data clips for pre-seizure
data segments and inter-ictal data clips for inter-seizure data segments. The corresponding segment length
is 10 minutes and the overall recording duration is 100 hours for each subject. The EEG signals of both
inter-ictal and pre-ictal events are simultaneously recorded using 16 electrodes the capture the same signal
from different neural channels. Both data categories are discretized with a sampling frequency is 400 Hz.
The pre-ictal data was recorded to cover one hour before seizure occurrence with 5 minutes seizure horizon
while the inter-ictal data was restrictedly recorded at least 4 hours before and after seizure.
\par
\begin{figure}[!htb]
    \centering
    \includegraphics[scale=1.05]{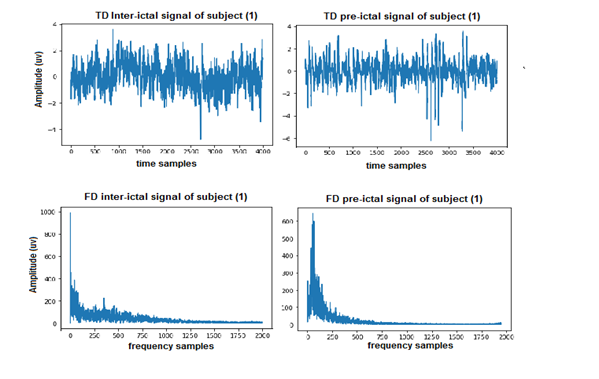}
    \caption{Inter-ictal vs. pre-ictal signal of the same subject}
    \label{fig4}
\end{figure}

\begin{figure}[!htb]
    \centering
    \includegraphics[scale=1.05]{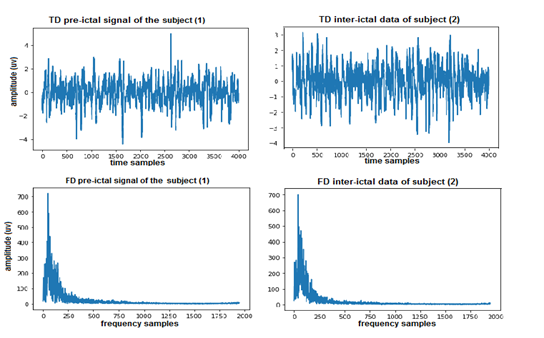}
    \caption{Inter-ictal vs. pre-ictal signals of different subjects}
    \label{fig5}
\end{figure}

 \hspace{0.3in}Initial inspection of the dataset raises many challenges that hardens the process of developing a
universal model to detect seizure events for non-patient-specific scenarios. One challenge include the fact
that pre-ictal data for one subject may appear similar to inter-ictal data of the same subject, both in time and
frequency domain. This can be exemplified in figure \ref{fig4}. Showing the time domain (TD) and frequency
domain of 10-second (i.e. 4000 samples) inter-ictal and pre-ictal signals, the two signals are hardly
distinguishable for the same subject. Another challenge involves that inter-ictal states of one subject may
show similar patterns to pre-ictal state of another subject for both time and frequency domain. This is shown
in figure \ref{fig5}.

\par \hspace{0.3in}

\begin{figure}[!htb]
    \centering
    \includegraphics[height=21cm,width=17cm]{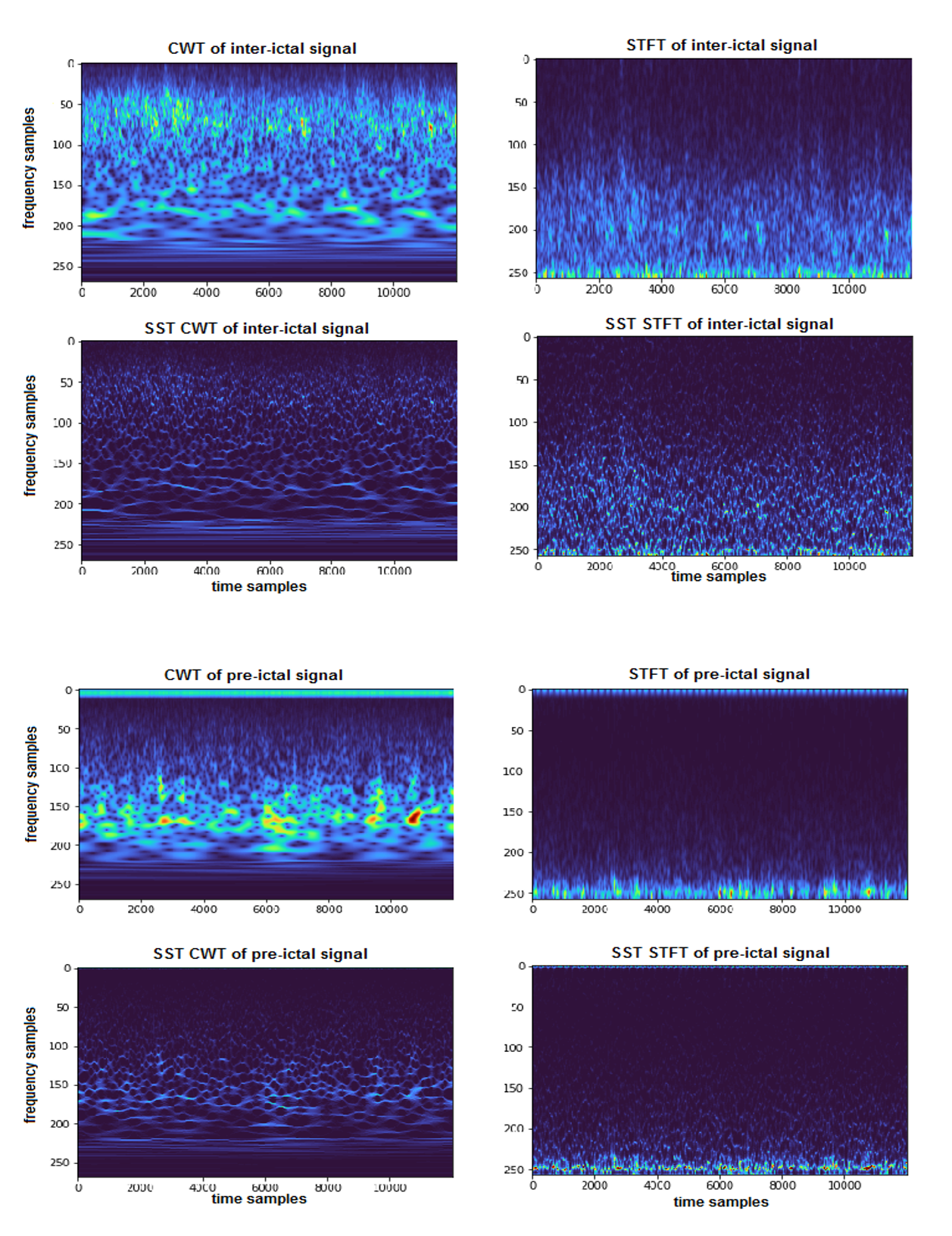}
    \caption{CWT and STFT TFRs and their SST versions for inter-ictal and pre-ictal signal}
    \label{fig6}
\end{figure}

\hspace{0.3in}Analysis of the EEG data segments in TF plane does not exhibit any apparent instantaneous
properties of the analyzed signal. Physical observations of both inter-ictal and pre-ictal data segments shows
random distributions of spectral components that cannot establish a robust basis for distinguishing between
inter-ictal and pre-ictal datasets. One way to demonstrate this by obtaining the TFRs, STFT and CWT and
their squeezed versions of 30-seconds (i.e. 12000 samples) inter-ictal and pre-ictal segments. As shown in
figure \ref{fig6}, the observed TF patterns do not show any anomalies that assist understanding the current state of
the segment. While TF unclarity persists, the SST versions of STFT and CWT for both inter-ictal and preictal data segments provide more concentrated representations which is anticipated to assist the CNN model
in extracting useful features from the inputted EEG TFRs. \par

\hspace{0.3in}Now that the data is segmented and its TFRs are obtained, the preprocessed data is split into
separate training and testing sets to be fed to the CNN prediction model. For this experiment, we only
consider the setting of STFT and its synchrosqueezed version for data modeling. We hereafter discuss the
relative performance of feeding SST TFRs to conventional STFT-based CNN prediction model.

\subsubsection{CNN prediction model description}
\begin{figure}[!htb]
    \centering
    \includegraphics[height=8cm,width=15cm]{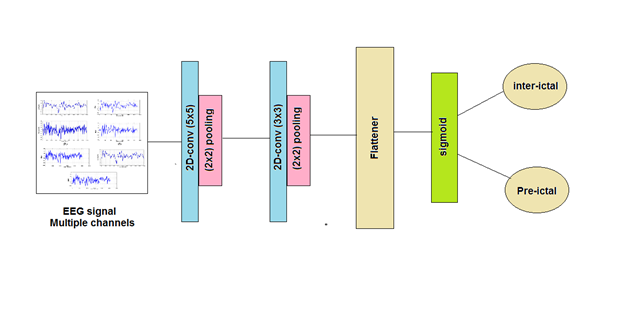}
    \caption{CNN-based prediction model components}
    \label{fig7}
\end{figure}

The prediction model is proposed by [8] which adopts two 2D convolutional layers, one with kernel
size of (5,5) and the other of size (3,3). The two layers extract abstract feature in a hierarchal fashion to
model temporal and spectral sequence information. In other words, The CNN layers performs in-depth
filtering to learn the weights associated to each channel (measurement of electrodes) to suitably integrate
the signal through the channels. The two convolutional layers are followed by pooling module of a kernel
size (2,2), flattening and sigmoid modules as shown in figure \ref{fig7}.

\subsubsection{Performance analysis}
\hspace{0.3in}The patient-based seizure prediction model is trained using EEG sample data related to individual
patient and the testing is performed on another patient data to ensure universal operation of the model. The
EEG time series mapping to TF plane is done in the order of 5000 samples at a time. The STFT, and
subsequently the STFT-based SST, are obtained by sliding a Slepian window with a hop length of 224 samples. The model is therefore trained twice and accuracies are compared. As anticipated, training the
model with synchrosqueezed samples results in improved accuracy in the order of 89.67% and a test loss
of 0.2653 compared to accuracy of 86.50% and test loss of 0.3161 of the STFT-trained model. Therefore,
the SST provides better assistance for extracting texture features out of nonlinear EEG signals resulting in
optimized performance.

\section{Results and Discussion}
\hspace{0.3in}The SST has proven significant usefulness in wide range of theoretical and practical applications. As
linear TFRs experience TF resolution tradeoff and manifest blurry representations of MCSs, components
detectability and hence invertibility is hindered by mode mixing problem. That is to say, SST is built on the
top of these linear TFRs and provides more focused representations while preserving the overall signal
energy. Regarding components retrieval, SST utilizes signal sparsity and formulates accurate formula with
variable entropy for reconstructing individual MCS components. \par
\hspace{0.3in}Relative comparison of STFT-based SST and CWT-based SST indicates that CWT-based SST is better
suited for analyzing low frequencies due to high frequency resolution at lower frequency values. On the other hand, STFT-based SST maintain constant resolution over all frequency range, making it a good fit for
reproducing MCS modes with closely packed instantaneous properties.\par
\hspace{0.3in}The effectiveness of SST can be further addressed through physical examples of analyzing signals of
varying oscillatory nature and/or embedded in severe noisy environments. For instance, SST can be
employed to preprocess time series data for DL-based classification/prediction models. Compared to
traditional TFRs preprocessing, SST proposes better approaches for feature extraction and anomalies
identification. The previously studied the epileptic seizure prediction system serves as a good example of
SST efficiency as SST-based prediction models outperform those of STFT by remarkably enhancing model
accuracy and reducing test losses.

\section{Conclusion}
\label{con}
\hspace{.3in}In this study, the SST theory is surveyed for both STFT-based and CWT-based approaches. Mode
retrieval operation is explained with corresponding applicability conditions. The advantageous role of SST
in high-resolution physical analysis is inspected using synthetic example as well as real-life practical
application involving CNN seizure prediction model. The performed experiments infer that SST aids better
extraction of features and improves DL model accuracies compared against conventional model relying on
linear TFRs. It is concluded that SST improves frequency resolution, allows precise component
reconstruction and enhances DL-based model accuracies when applied to preprocess time series signals.

\bibliographystyle{unsrt}
\bibliography{references}

\begin{thebibliography}{1}

\bibitem{thakur2015synchrosqueezing}
Gaurav Thakur.
\newblock The synchrosqueezing transform for instantaneous spectral analysis.
\newblock In {\em Excursions in Harmonic Analysis, Volume 4}, pages 397--406.
  Springer, 2015.

\bibitem{li2012generalized}
Chuan Li and Ming Liang.
\newblock A generalized synchrosqueezing transform for enhancing signal
  time--frequency representation.
\newblock {\em Signal Processing}, 92(9):2264--2274, 2012.

\bibitem{meignen2019synchrosqueezing}
Sylvain Meignen, Thomas Oberlin, and Duong-Hung Pham.
\newblock Synchrosqueezing transforms: From low-to high-frequency modulations
  and perspectives.
\newblock {\em Comptes Rendus Physique}, 20(5):449--460, 2019.

\bibitem{muradeli_2022}
John Muradeli.
\newblock Overlordgolddragon/ssqueezepy: Qol, cleanups, fixes, Jan 2022.

\bibitem{hussein2020augmenting}
Amir Hussein, Marc Djandji, Reem~A Mahmoud, Mohamad Dhaybi, and Hazem Hajj.
\newblock Augmenting dl with adversarial training for robust prediction of
  epilepsy seizures.
\newblock {\em ACM Transactions on Computing for Healthcare}, 1(3):1--18, 2020.

\bibitem{kaggle}
American epilepsy society seizure prediction challenge.

\end{thebibliography}

\end{document}